\documentclass[aps,prb,twocolumn,showpacs]{revtex4}
\usepackage{amssymb}

\usepackage{graphicx}

\begin{document}

\title{Power Law Like Correlation between Condensation Energy and Superconducting Transition Temperatures in Iron Pnictide/Chalcogenide Superconductors: Beyond the BCS Understanding}

\author{Jie Xing$^1$, Sheng Li$^1$, Bin Zeng$^2$, Gang Mu$^2$, Bing Shen$^2$, J. Schneeloch$^3$, R. D. Zhong$^3$, T. S. Liu$^{3,4}$, G. D. Gu$^{3}$, and Hai-Hu Wen$^1$ }\email{hhwen@nju.edu.cn}

\affiliation{$^1$Center for Superconducting Physics and Materials, National Laboratory of Solid State Microstructures and Department of Physics, Nanjing University, Nanjing 210093, China}

\affiliation{$^{2}$ National Laboratory for Superconductivity,
Institute of Physics and National Laboratory for Condensed Matter
Physics, Chinese Academy of Sciences, Beijing 100190, China}

\affiliation{$^{3}$ Condensed Matter Physics and Materials Science Department,
Brookhaven National Laboratory, Upton, NY 11973, USA}

\affiliation{$^{4}$ School of Chemical Engineering and Environment, North University of China, Shanxi 030051, China}

\date{\today}

\begin{abstract}
Superconducting condensation energy $U_0^{int}$ has been determined by integrating the electronic entropy in various iron pnictide/chalcogenide superconducting systems. It is found that $U_0^{int}\propto T_c^n$ with $n$ = 3 to 4, which is in sharp contrast to the simple BCS prediction $U_0^{BCS}=1/2N_F\Delta_s^2$ with $N_F$ the quasiparticle density of states at the Fermi energy, $\Delta_s$ the superconducting gap. A similar correlation holds if we compute the condensation energy through $U_0^{cal}=3\gamma_n^{eff}\Delta_s^2/4\pi^2k_B^2$ with $\gamma_n^{eff}$ the effective normal state electronic specific heat coefficient. This indicates a general relationship $\gamma_n^{eff} \propto T_c^m$ with $m$ = 1 to 2, which is not predicted by the BCS scheme. A picture based on quantum criticality is proposed to explain this phenomenon. \end{abstract}


\maketitle
Superconductivity is induced by quantum condensation of large
number of paired electrons, namely the Cooper pairs. According to the
Bardeen-Cooper-Schrieffer (BCS) theory, the pairing is supposed to be
established between the two electrons with opposite momentum and
spins by exchanging phonons. The formation of the electronic paired state will lower the total energy leading to the condensation of the Cooper pairs. The condensation energy, defined as the the difference of the Gibbs free energy of the system in the normal state and superconducting state, is given by $U_0^{BCS}=1/2N_F\Delta_s^2$ with $N_F$ the quasi-particle density of states (DOS) at the Fermi energy of the normal state, $\Delta_s$ is the superconducting gap. Suppose a simple and natural relation, $\Delta_s \propto T_c$, we have $U_0^{BCS}\propto N_F T_c^2$. Normally $N_F$ is weakly related to the superconducting gap $\Delta_s$ through $N_F=1/Vln[(2\hbar \omega_{D})/\Delta_s]$ with $V$ the attractive potential between the two electrons when exchanging a phonon and $\omega_D$ the Debye frequency, thus one can roughly expect that $U_0^{BCS}\propto T_c^2$ in a conventional BCS superconductor.

Since the discovery of iron based superconductors, the pairing
mechanism remains unresolved yet. One type of picture assumes the
similar scenario of the BCS but using the antifferomagnetic spin
fluctuations as the pairing
glue\cite{MazinPRL2008,Kuroki2008,Hirschfeld2008,LeeDH2009}. This is
called the weak coupling approach. Another more exotic picture,
based on the strong coupling approach, assumes the local magnetic
interaction as the pairing force which simultaneously causes the
pairing of two electrons\cite{HuJP,SiQM,Zaanen,Kotliar}. However,
both pictures will intimately lead to an $s^\pm$ pairing gap as the
natural one. Specific heat (SH) measurements are very powerful, not
only in detecting the gap symmetry\cite{MuGPRB2008,HardyEPL,Keimer},
but also in unraveling some deeper mysteries related to the superconducting
mechanism. For example, it was found by Bud'ko, Ni and Canfield (BNC)\cite{BNC} that, in the 122 systems, there is a simple scaling relation $\Delta C|_{T_c} \propto
T_c^3$ with $\Delta C|_{T_c}$ the SH anomaly (jump) at T$_c$. This
simple relation was later proved and solidified by further
measurements with the samples experienced different thermal treatments and
annealing\cite{Stewart}, and extended to the 11 and 111 systems also\cite{Stewart3,BNC2}. This $\Delta C|_{T_c} \propto T_c^3$
relation was explained as due to the impurity scattering effect in a
multiband superconductor with the $s^{\pm}$ pairing gap\cite{Kogan}.
However, this explanation may suffer a challenge when making a
comparison between Ba$_{1-x}$K$_x$Fe$_2$As$_2$ and
Ba(Fe$_{1-x}$Co$_x$)$_2$As$_2$: the former is much cleaner than the
latter judged through the residual scattering
rate\cite{ShenBPRB2011}, but they follow a similar trend in the
scaling relation $\Delta C|_{T_c} \propto T_c^3$. Another more novel
picture, concerned with the quantum critical point (QCP)\cite{Zaanen2}, was proposed to
understand this interesting relation. Since the
condensation energy is directly related to how much energy that is
saved when the system enters the superconducting state, thus it is highly desired to have a systematic
assessment on the condensation energy. In
this Letter we try to calculate the condensation energy from 10
pieces of our measured single crystals, and others from the
published literatures. Surprisingly we discovered a simple power law
like relation between the condensation energy and the superconducting
transition temperatures.

Single crystals of Ba$_{1-x}$K$_x$Fe$_2$As$_2$ (x=0.3,0.4),
Ba(Fe$_{1-x}$Co$_x$)$_2$As$_2$, BaFe$_{1.9}$Ni$_{0.1}$As$_2$ were
grown by the flux method\cite{MuGPRB2008,OurBaCo}, the
FeSe$_{0.5}$Te$_{0.5}$ by a unidirectional solidification
method\cite{GDGu}. The SH data of
Ba$_{1-x}$K$_x$Fe$_2$As$_2$(x=0.4), Ba(Fe$_{1-x}$Co$_x$)$_2$As$_2$
and BaFe$_{1.9}$Ni$_{0.1}$As$_2$ single crystals were published in
previous papers\cite{MuGPRB2008,OurBaCo}. All doping concentrations
of our samples are the nominal ones. The SH measurements
were done by the thermal relaxation method on the physical property
measurement system (PPMS, Quantum Design) with the advanced
measuring chip. For determining the condensation energy, we properly removed the phonon contributions (see below). We also get the electronic SH data from the
published papers of other
groups\cite{FeSeTeMao,FeSeTeTsurkan,FeSeTeKlein,FeSeTeGunther,FeSeLin,LiFeAsStockert,LiFeAsWei,BaKPopovich,BaKWei,BaCoGofryk,BaNaPramanik,FeSeTeNoji,BaCoGofrykPRB}
so as to make the statistic results more convincing.

\begin{figure}
\includegraphics[width=8cm]{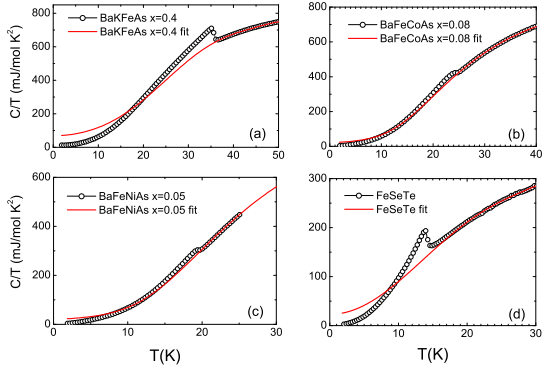}
\caption {(color online) Raw data of SH for four
different superconducting systems near the optimal doping point. The
data are shown for samples (a) Ba$_{0.6}$K$_{0.4}$Fe$_2$As$_2$, (b)
Ba(Fe$_{0.92}$Co$_{0.08}$)$_2$As$_2$; (c) BaFe$_{1.9}$Ni$_{0.1}$As$_2$;
and (d) FeSe$_{0.5}$Te$_{0.5}$. Here we show only four typical sets
of data and the fitting curves of the normal state. More data are
presented in the Supplementary Materials.} \label{fig1}
\end{figure}

In Fig.~1, we present the temperature dependence of SH
for four typical samples from the ten. The sharp SH anomaly
can be seen clearly at T$_c$ for each sample. In order to obtain the
electronic SH, we have to investigate the phonon part of
the total SH carefully. For
Ba(Fe$_{1-x}$Co$_x$)$_2$As$_2$ and BaFe$_{1.9}$Ni$_{0.1}$As$_2$
samples, because the phonon contribution changes not much with
doping, the overdoped nonsuperconducting samples are used as the
references. Thus, from the formula
\begin{equation}
C_{e}^{s}(T)=C_{total}^{s}(T)-p*C_{ph}^{n}(q*T)
\end{equation}
we can derive the electronic term in each superconducting sample.
Here C$_{e}^{s}(T)$, C$_{total}^{s}$(T) are the electronic and total
SH of the superconducting samples respectively,
C$_{ph}^{n}$(T) is the phonon contribution of SH of the reference one. The $p$
and $q$ are fitting parameters which are determined by having a
close matching effect of the phonon part between the superconducting
sample and the reference one.  It is found that $p$ and $q$ are
close to 1.\cite{OurBaCo} This slight modification of the
phonon contribution is understandable since the doping may change
the lattice constants slightly. For Ba$_{1-x}$K$_x$Fe$_2$As$_2$ and
FeSe$_{0.5}$Te$_{0.5}$ samples, we use a polynomial function
C$_{nor}$=C$_{e}$+C$_{ph}$=
$\alpha$T+$\beta$T$^3$+$\gamma$T$^5$+$\cdot$$\cdot$$\cdot$ to fit
the data in the normal state above T$_c$. In the fitting process, to
ensure the entropy conservation, we leave the electronic term
$\alpha$ as the trying parameter and leave other higher-power
temperature related terms totally free. The red
lines in Fig.~1 show the phonon and the normal state electronic contribution of each sample. Either
using a reference sample or using the polynomial fitting method, one
can find a good fit of the normal state of each superconducting
sample. We must emphasize that, to ensure the the entropy
conservation is a basic rule we hold in removing the phonon
contribution in either methods mentioned above. This may inevitably
lead to some uncertainties of the condensation energy with the error
bars of about $\pm$10\%. For clarity, we only show data for four optimally doped
samples in Fig.~1 and the data of other six samples are presented in
the Supplementary Materials (SM).

\begin{figure}
\includegraphics[width=8cm]{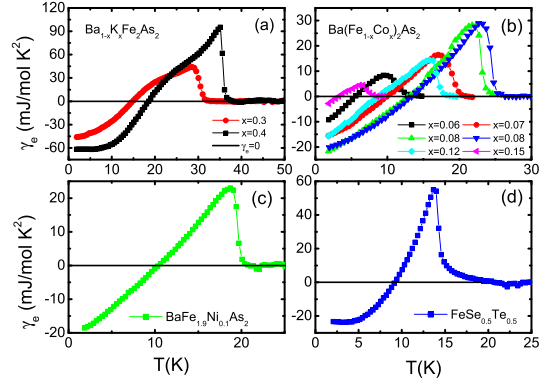}
\caption {(color online) Temperature dependence of the superconducting electronic specific heat shifted by $\gamma_n^{eff}+\gamma_0$, i. e., $C_e/T-\gamma_n^{eff}-\gamma_0$ for (a)
Ba$_{1-x}$K$_x$Fe$_2$As$_2$, (b) Ba(Fe$_{1-x}$Co$_x$)$_2$As$_2$; (c)
BaFe$_{1.9}$Ni$_{0.1}$As$_2$; and (d) FeSe$_{0.5}$Te$_{0.5}$.
}\label{fig2}
\end{figure}

After subtracting the phonon contribution from the total SH, the electronic contribution is obtained for our ten samples,
as shown in Fig.~2. The residual term at $T=0$ K gives actually the effective
SH coefficient -$\gamma_n^{eff}=-(\gamma_{n}-\gamma(0)$), with
$\gamma_n$ the total electronic SH of the normal state,
including the nonsuperconducting term $\gamma_0$.\cite{Nodal} The
SH anomaly at T$_c$ rises to a maximum at optimal doping
point with the highest T$_c$. Above T$_c$, the electronic SH decreases rapidly except for FeSe$_{0.5}$Te$_{0.5}$. For
FeSe$_{0.5}$Te$_{0.5}$, there is a tail extending up to a higher
temperature, which may suggest that this system can be made with
higher transition temperatures, as achieved in thin films\cite{LiQ}.
This phenomenon was found by other groups as
well\cite{FeSeTeMao,FeSeTeTsurkan,FeSeTeKlein,FeSeTeGunther}.

According to the BCS theory, the SH anomaly of a
superconductor at T$_c$ should follow $\Delta$C/$\gamma$$_n$T$_c$ =
1.43 in the weak coupling limit. However, it was found that the iron
based superconductors severally violate this relation but show a
simple correlation $\Delta C|_{T_c} \propto T_c^3$. This power law
seems to be appropriate for many iron based superconductors, with the majority of data so far for the 122 systems\cite{BNC,Stewart3,BNC2}. We also determined the SH anomaly
of our ten samples and show them together with those of BNC in
Fig.~3. Because of the finite width of the superconducting
transition at T$_c$, we use the entropy conservation to determine the
value of SH anomaly and T$_c$ in our samples. It's clear that our data fall onto the general power law  $\Delta C|_{T_c}
\propto T_c^3$ quite well except for the FeSe$_{0.5}$Te$_{0.5}$
sample on which a deviation is observed. This may suggest that the
general scaling law works better for one system, for example, for
122 here. However, we will show later that a scaling law of
condensation energy with T$_c$ seems more general to cover data from
different systems, such as 122, 111 and 11.

\begin{figure}
\includegraphics[width=8cm]{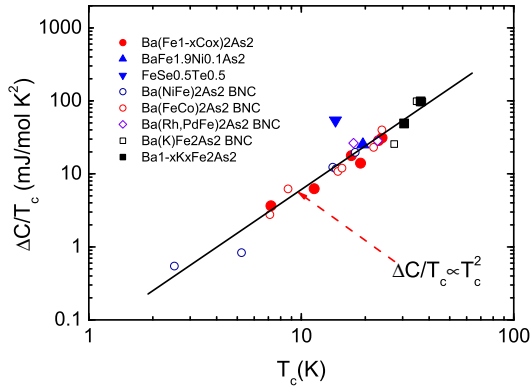}
\caption {(color online) Correlations between the SH
anomaly at $T_c$, i.e., $\Delta C/T_c|_{T_c}$ and $T_c$ for many
iron based superconductors. The solid line shows the relationship with
$\Delta C/T_c|_{T_c} \propto T_c^2$. The solid symbols are from our present experiment. The open ones are from the work of BNC.}\label{fig3}
\end{figure}

In addition to the SH anomaly, the condensation energy is
another important parameter to determine the properties underlying
the superconductivity. According to the
thermodynamic definition, the entropy is $S=-\partial G/\partial T$,
therefore we can calculate the condensation energy by
integrating the entropy of the superconducting and normal state,
\begin{equation}
U_0^{int}=\int_{0}^{T_c}(S_n(T)-S_s(T))dT
\end{equation}
\begin{equation}
=\int_{0}^{T_c}dT\int_{0}^T(C_n(T')-C_s(T'))/T'dT'.
\end{equation}

The temperature dependence of entropy is shown in the SM
for the four optimally doped samples. Because
($C_n(T)-C_s(T))/T$=$\gamma_n(T)$-$\gamma_s(T)$, we can just
compute condensation energy with the electronic SH. We
also calculate the condensation energy using the electronic SH data in previously
published
papers\cite{FeSeTeMao,FeSeTeTsurkan,FeSeLin,LiFeAsStockert,LiFeAsWei,BaKPopovich,BaKWei,BaCoJang,BaCoGofryk,BaNaPramanik,FeSeTeNoji,BaCoGofrykPRB,BaKpoly}. These data are plotted together with ours in Fig.~4(a). The dashed line shows the correlation
$U_0^{int}\propto T_c^{3.5}$. For different systems, the exponent
$n$ may vary a little bit, for example for the
Ba$_{1-x}$K$_x$Fe$_2$As$_2$, $n$ is slightly smaller than that in
Ba(Fe$_{1-x}$T$_x$)$_2$As$_2$ (T=Co and Ni). However, a global scaling law can be
roughly satisfied with the exponent $n$ $\approx$ 3-4. Because the
fermionic DOS should be weakly dependent on the doping level across the
optimally doped point, the BCS theory implies that the condensation
energy should scale roughly with $T_c^2$, which is very different
from our result. We should mention that some published results from samples (mostly in the 111 system) with broad superconducting transitions are not included here. It is thus very curious to know whether more data points from variety of
systems are also obeying this scaling law. Furthermore, the SH data from the K$_x$Fe$_{2-y}$Se$_2$ and
KFe$_2$As$_2$ systems are not included. This is justified by the phase separation\cite{WenHHReview} in K$_x$Fe$_{2-y}$Se$_2$. For the
KFe$_2$As$_2$ system, the $T_c$ is too low, which may prevent determining the condensation energy precisely\cite{Stewart2}.

Taking account of the BCS theory, we
can deduce the condensation energy from the known values of
$\gamma_n^{eff}$ and the gap $\Delta_s$ as well. As a first approximation, assuming a spherical Fermi surface, the condensation energy is given by
$U_0^{cal}=1/2N_F\Delta_s ^2$ with the DOS
$N_F=3\gamma_n^{eff}/(2\pi^2k_B^2$) with $\gamma_n^{eff}=\gamma_n-\gamma_0$. From this argument,
the condensation energy is derived as

\begin{equation}
U_0^{cal}=\frac{3(\gamma_n^{eff})}{4\pi^2k_B^2}\Delta_s^2.
\end{equation}

Starting from above equation and the values of $\gamma_n^{eff}$ and the gap, we calculate the
condensation energy in an alternative way for our four optimally doped samples on which both the $\gamma_n^{eff}$ and $\Delta_s$ are available, and from the published
data for other samples\cite{MuGPRB2008,OurBaCo,FeSeTeMao,FeSeTeTsurkan,FeSeLin,LiFeAsStockert,LiFeAsWei,BaKPopovich,BaKWei,BaCoGofryk,BaNaPramanik,FeSeTeNoji,BaCoGofrykPRB}.
Because of the multigap feature in the iron pnictide
superconductors, some samples were fit by two s-wave gaps so
we used the average gap
$\Delta_s$=$\sqrt{((p_1\Delta_1)^2+(p_2\Delta_2)^2}$. For a d-wave
component, the effective gap $\Delta_s$=$\frac{\sqrt2}{2}\Delta_d$ (here $\Delta_d$ is the maxima of the d-wave gap) is
used in the formula. The calculated data of condensation
energy are plotted in Fig.~4(b). The dashed line shows the power law
$U_0^{cal}\propto T_c^{3.5}$. To our surprise, not only $U_0^{int}$,
but also the calculated value of the condensation energy $U_0^{cal}$
also obeys the correlation $U_0\propto T_c^n$ with $n$ of about 3-4. The result
strongly indicates that the correlation between condensation energy
and T$_c$ reveals the intrinsic property in iron based
superconductors. If we look back to the BNC relation, $\Delta
C/T_c|_{T_c}$ $\propto$ $T_c^2$, a slight difference between our
result and BNC relation can be found by using the BCS theory. Taking
$U_0^{BCS}$ = $1/2N_F\Delta_s^2$, $\Delta_s$ = 1.75 $k_BT_c$,
$N_F=3\gamma_n^{eff}$/(2$\pi^2$k$_B^2$), we have
$\Delta C|_{T_c}$ = 1.43$\gamma_n^{eff} T_c$= 6.14$U_0^{BCS}$/T$_c$. This would suggest from
our result that $\Delta C|_{T_c}\propto T_c^{2.5}$. This discrepancy
 further suggests that the simple BCS formulas, especially those based on the weak coupling approach, cannot be used in the iron
based superconductors. Nevertheless, either the power law like relation found by BNC about the $\Delta C|_{T_c}$ vs. $T_c$, or that between the condensation energy and $T_c$, are beyond the expectations by the BCS theory. In the following, we argue that the doping dependence of the effective DOS (or $\gamma_n^{eff}$) may play an important role here.

\begin{figure}
\includegraphics[width=8cm]{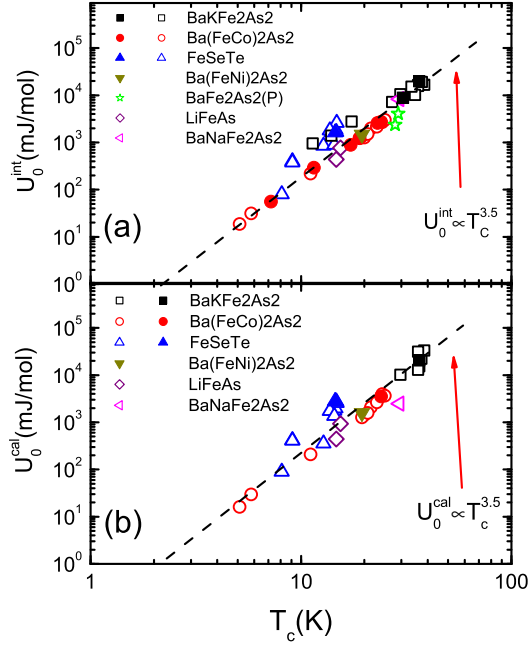}
\caption {(color online)Correlations between the condensation energy
and $T_c$ in several iron
based systems. Here the condensation energy is
calculated through (a) integrating the entropy in the
superconducting state (see text) and (b) the simple computing
formula $U_0^{BCS}$ = $1/2N_F\Delta_s^2$. The dashed lines
represent the relation $U_0^{int}$ or $U_0^{cal}\propto T_c^{3.5}$. Here the filled symbols are from our experiment, the open ones are from the available literatures. }
\label{fig4}
\end{figure}

Now we investigate the doping dependence of the condensation energy and
the effective SH coefficient $\gamma_n^{eff}$ in 122 system. The results are shown in Fig.~5. It contains not only the data of our 9 samples, but also some available data from literatures. The x-coordinate is the doped
charges per Fe for every compound. In both doping sides, the quantities $U_0^{int}/T_c^2$, $U_0^{cal}/T_c^2$ and $\gamma_n^{eff}$
overlap quite well and all exhibit a maximum around the optimal doping point. Taking account the result $U_0 \propto T_c^n$ with $n$ = 3-4, we have $\gamma_n^{eff} \propto T_c^{m}$ with $m$ = 1-2. This is not expected by the BCS theory. Since $\gamma_n^{eff}$ is closely related to the effective mass, we intend to argue that this novel doping dependence of $\gamma_n^{eff}$ (or the effective DOS) results from the mass enhancement when it is around the quantum critical point (QCP).

\begin{figure}
\includegraphics[width=8cm]{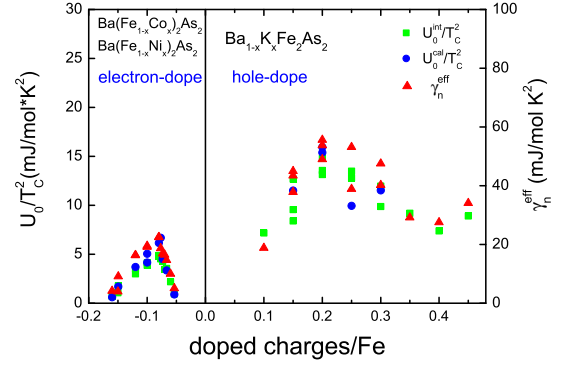}
\caption {(color online)Doping dependence of $U_0^{int}/T_c^2$, $U_0^{cal}/T_c^2$ and $\gamma_n^{eff}$ for electron doped and hole doped 122 samples. The three set of data overlap each other. } \label{fig5}
\end{figure}

As we know, the antiferromagnetism and superconductivity appear closely in the electronic phase diagram revealed either by doping or by applying a high pressure in iron based superconductors. In most systems, if extrapolating the antiferromagnetic (AF) transition to zero temperature, it is found that the highest T$_c$ appears near the point where the Neel temperature of the AF order becomes zero and a strong AF spin fluctuation emerges\cite{NingFL}. Near the optimal doping point, many novel electronic properties have been observed, for example the penetration depth seems to have a singularity in P-doped BaFe$_2$As$_2$ system\cite{QCPMatsuda}. Therefore it is quite possible that the effective mass of the electrons are strongly enhanced due to the strong coupling between the electrons and the AF spin fluctuations. This possible effect may bring about the power law like correlation between the condensation energy and T$_c$. It was also discovered that the enhancement of effective mass appears near the quantum critical point in cuprates\cite{QCPCuprate}. The divergent effective mass was found in the heavy fermion system near the antiferromagnetic quantum critical point as well\cite{HF}. Another feasible explanation which may be related to the above mentioned QCP mechanism is the small Fermi energy $E_F$ in many iron-based superconductors. In the usual situation for the BCS picture, it is known that $\omega_{D}$/E$_F$$\ll$1, in this case the pair-scattering occurs only near the very thin shell of the Fermi surface. While in the iron-based superconductors, there are many shallow bands crossing the Fermi level leading to a small Fermi energy $E_F$. This may further enhance the quantum fluctuation effect of the electronic system. Our observation here, that is $U_0 \propto T_c^n$ with $n$ = 3-4, can be explained as a consequence of the QCP as argued by Zaanen\cite{Zaanen2}. This will certainly stimulate further theoretical and experimental efforts on this general and interesting phenomenon.

In conclusion, the SH of many iron based superconductors in the 122, 11 and 111 systems was investigated. From these data, we computed the condensation energy by two different methods and get similar power law like correlations $U_0^{int}\propto T_c^{n}$ and $U_0^{cal}\propto T_c^{n}$ with $n$ = 3-4. Combining this relationship and the semi-quantitative consideration of the BCS theory $U_0^{BCS}$ = $1/2N_F\Delta_s^2$, we find that the effective SH coefficient $\gamma_n^{eff}$, or the effective DOS is proportional to T$_c^{m}$ with $m$ = 1-2 across the doping regime, either in the electron or the hole doping side. All these power law like relations are beyond the BCS understanding, but can be explained based on the QCP picture. This discovery reveals the originality that is intimately related to the unconventional superconducting mechanism.

\begin{acknowledgments}
We appreciate the useful discussions with Tao Xiang and Qimiao Si. This work is supported by the
Ministry of Science and Technology of China (973 Projects: No.
2011CBA00102, No. 2010CB923002, and No. 2012CB821403), the NSF of
China, NCET project and PAPD. Work at Brookhaven was supported by DOE through contract No. DE-AC02-98CH10886.

\end{acknowledgments}


\begin{thebibliography}{99}
\bibitem{Nodal} For a superconductor with nodal gaps, there is a residual term due to the impurity scattering effect. This should not be the case in our present samples.
\end{thebibliography}
\end{document}